\documentclass[twocolumn,superscriptaddress,nofootinbib,notitlepage,longbibliography,aps]{revtex4-1}
\pdfoutput=1

\usepackage{dcolumn}
\usepackage[pdftex]{graphicx}
\usepackage{bm}
\usepackage{array}
\usepackage[utf8]{inputenc}
\usepackage{amsmath}
\usepackage{amsfonts}
\usepackage{bbold}
\usepackage{graphicx}
\usepackage{float}
\usepackage{braket}
\usepackage[T1]{fontenc}
\usepackage[english]{babel}
\usepackage{amsthm}
\usepackage{makeidx}
\usepackage{amssymb}

\usepackage{braket}
\usepackage{mathtools}
\usepackage{caption}
\usepackage[english]{varioref}
\usepackage{url}
\usepackage[pagebackref]{hyperref}
\usepackage{float}
\restylefloat{table}
\usepackage{multirow}
\usepackage{xcolor}

\usepackage{dcolumn}

\usepackage{bm}
\usepackage{array}
\usepackage[utf8]{inputenc}
\usepackage{amsmath}
\usepackage{amsfonts}
\usepackage{bbold}
\usepackage{graphicx}
\usepackage{float}
\usepackage{braket}
\usepackage{slashed}
\usepackage[T1]{fontenc}
\usepackage[english]{babel}
\usepackage{amsthm}
\usepackage{makeidx}
\usepackage{amssymb}
\usepackage{mathtools}
\usepackage{caption}
\usepackage[english]{varioref}
\usepackage{url}
\usepackage{float}
\restylefloat{table}
\usepackage{multirow}
\usepackage{xcolor}
\usepackage{appendix}

\newcommand{\INFN}{INFN - Sezione di Napoli, Complesso Universitario Monte S. Angelo, I-80126 Napoli, Italy}
\newcommand{\UNINA}{Dipartimento di Fisica ``Ettore Pancini'', Università degli studi di Napoli ``Federico II'', Complesso Universitario Monte S. Angelo, I-80126 Napoli, Italy}
\newcommand{\SSM}{Scuola Superiore Meridionale, Università degli studi di Napoli ``Federico II'', Largo San Marcellino 10, 80138 Napoli, Italy}

\begin{document}
\title{ \boldmath Towards baryogenesis via absorption from Primordial Black Holes 
%
}
\author{Antonio Ambrosone}
\email{aambrosone@na.infn.it}
\affiliation{\UNINA}
\affiliation{\INFN}

\author{Roberta Calabrese}
\email{rcalabrese@na.infn.it}
\affiliation{\UNINA}
\affiliation{\INFN}
\author{Damiano F.G. Fiorillo}
\email{dfgfiorillo@na.infn.it}
\affiliation{\UNINA}
\affiliation{\INFN}

\author{Gennaro Miele,}
\email{miele@na.infn.it}
\affiliation{\UNINA}
\affiliation{\INFN}
\affiliation{\SSM}

\author{and Stefano Morisi}%
\email{smorisi@na.infn.it}
\affiliation{\UNINA}
\affiliation{\INFN}

\begin{abstract}

Recently Dolgov and Pozdnyakov proposed a new baryogenesis mechanism
in which baryon asymmetry is produced
without violating baryon number at the Lagrangian level. In this scenario,
baryon asymmetry is generated by absorption of a new particle X carrying
baryon number onto Primordial Black Holes (PBHs). Assuming CP-violation, the
particle X is absorbed at a different rate than the antiparticle $\bar{X}$, producing an asymmetry in the
baryonic number. We independently test this scenario, finding that it suffers from two fundamental issues.\\
At the phenomenological level, strong absorption by PBHs initially increases the baryon asymmetry. However, at later times such asymmetry is completely absorbed by PBHs. In order to overcome this issue, we account for PBH evaporation, which provides a natural way of halting the absorption while keeping a finite baryon asymmetry. We provide a systematic study of the parameter space,
identifying the regions leading to the production of the baryon asymmetry
without violating the known constraints on PBHs concentration. 

At the theoretical level, a model realizing the CP-violation postulated in this scenario is difficult to realize.  We show, by implementing a minimal model, that the framework proposed in the original work in order to produce CP-violation, even if qualitatively correct, is quantitatively in disagreement with the observed baryon asymmetry, namely this
mechanism produces only a fraction of the total baryon asymmetry. 
\end{abstract}

\maketitle

\section{Introduction}
Our universe exhibits an asymmetry among baryons and anti-baryons, whose origin is one of the fundamental problems of physics. The amount of baryon asymmetry can be quantified by the measured ratio~\cite{ Davidson_2008,Copi:1994ev, Buchmuller:2002rq, aghanim2020planck} $\eta^{\text{exp}}_B \equiv (n_B-n_{\bar B})/s=(8.718\pm 0.004)\cdot 10^{-11}$, where $n_B\,(n_{\bar B})$ is the baryon (anti-baryon) number density and $s$ is the entropy density. In 1967, Sakharov proposed three conditions~\cite{sakharov1998violation} necessary for a viable baryogenesis mechanism: i) violation of the baryon number, ii) violation of CP and C symmetries, and iii) out of equilibrium processes. While the Standard Model satisfies Sakharov conditions, it is not able to reproduce quantitatively the required asymmetry~\cite{Bochkarev:1987wf, Kajantie_1996, GAVELA_1994, Huet_1995, Gavela_1994_part2}. Therefore, a number of extensions have been proposed to reproduce the correct asymmetry, including electroweak baryogenesis~\cite{Cohen:1993nk,Rubakov:1996vz, Morrissey_2012}, baryogenesis via leptogenesis~\cite{fukugita1986barygenesis,Luty:1992un, Barbieri:1999ma, Chianese:2018rnq}, GUT baryogenesis~\cite{Riotto:1999yt}, and the Affleck-Dine mechanism~\cite{AFFLECK1985361}.

In a novel framework proposed in Refs.~\cite{Bambi_2009,Dolgov:2020kqj, Dolgov:2021tsa}, baryon asymmetry is produced by the different absorption of particles and antiparticles of a new species $X$ (assumed to be scalar) with mass $m_X$, carrying baryon number $B_X$, onto a population of primordial black holes (PBH)\,\cite{Dolgov:2019ncq, Khlopov_2010}.  Their coupling with Standard Model (hereafter denoted by SM) is indicated with $f$. These particles are attracted by gravity onto a population of PBHs with mass $M$. Assuming that $X$ and $\overline{X}$ have slightly different cross-sections with the SM plasma surrounding the PBHs due to a small C and CP violation, the absorption rates of $X$ and $\overline{X}$ by PBHs are different. Such an asymmetric absorption leads to a continuous production of asymmetry in the $X$ sector.  In Ref.~\cite{Dolgov:2020kqj} it is suggested that strong absorption on PBHs ($f\ll 1 $) is required to reproduce a large asymmetry. We have reanalyzed this low-coupling scenario, finding that it would lead to the complete absorption of $X$ and $\overline{X}$, giving a total  suppression of the baryon asymmetry. Therefore, the quantitative value of the observed baryon asymmetry is not reproducible in the scenario of Ref.~\cite{Dolgov:2020kqj}. Here we show that such a problem can be solved by taking into account the evaporation of PBHs  by means of Hawking radiation~\cite{hawking1974black,hawking1975particle, zeldovich1976charge, carr1976some}.
Evaporating PBHs have been well-studied in connection with baryon asymmetry 
generation~\cite{Turner:1979bt, turner1979baryon,Baumann:2007yr, Fujita:2014hha, Hamada:2016jnq, chaudhuri2020pbh,  Baldes_2020, Hooper_2021, das2021low,datta2020baryogenesis,Perez-Gonzalez:2020vnz} as well as with the dark matter problem~\cite{Baldes_2020,Carr:1976zz,masina2021dark} (see also \cite{Carr_2016,Morrison_2019,Carr_2020,Hertzberg:2020kpm,Green_2021,tashiro2021constraining, bernal2021gravitational,Auffinger_2021}).
PBHs are subjected to Hawking evaporation, emitting particles and gradually losing their mass. The evaporation rate determines the temperature at which PBHs are expected to evaporate~\cite{Carr:2009jm}
\begin{equation}
    T_{\text{e}} \simeq 10^{10} \, \text{GeV} \left( M_g \right)^{-3/2}.
\end{equation}
We denote by $M_g$ the mass of PBH expressed in grams, whereas by $M$, already introduced above, we denote the same mass in natural units. For the rest of this work, we will adopt natural units unless otherwise stated. For simplicity, we will model the PBH mass to be non-evolving for temperatures larger than $T_{\text{e}}$ and to disappear at lower temperatures. 
The same approximation is also adopted in Ref.~\cite{Baumann:2007yr}.
For PBH masses lighter than $10^9$ g, which is the range of interest for us, PBHs completely evaporate before the Big Bang Nucleosynthesis. The finite lifetime of PBHs due to evaporation provides a natural way of stopping the absorption before the species is completely depleted, thereby maintaining the asymmetry produced. 
A connected problem is that the remnant $X$ particles, being very massive, might dominate over radiation. In the mechanism we propose here, this problem is solved by allowing the $X$ species to decay into baryons, accordingly transferring the asymmetry from the $X$ sector to the baryonic one. We show that this phenomenological mechanism can produce a non-vanishing baryon asymmetry by explicitly solving the cosmological transport equations. Therefore, by accounting for PBH evaporation, we are able to avoid the issue of the complete absorption of the species.

However, we find a more serious issue in the possibility of realizing this scenario in a complete model. Indeed, the feasibility of the model is based on the phenomenological parameterization of the CP violation efficiency in the cross-sections of $X$ and $\bar{X}$ proposed in Refs.~\cite{Dolgov:2020kqj,Dolgov:2021tsa}. However, no quantitative model is provided which supports such a phenomenological parameterization. Here we show that the construction of such a model is far from trivial. We provide an explicit minimal model able to produce a non-vanishing baryon asymmetry. In doing so, we perform the first calculation in the literature, to our knowledge, of the CP violation in the scattering cross-sections of a particle and antiparticle. We find that the constraints imposed by the model are too stringent to account for the whole of the observed baryon asymmetry. Therefore, a theoretical model supporting this scenario as a feasible one for baryogenesis remains to be constructed, possibly by relieving some of the assumptions in the minimal model presented here.

We structure the discussion by first providing a phenomenological overview of the mechanism in Sec.~\ref{sec:mechanism}, where we estimate the parameters and obtain the equations for the cosmological evolution of the baryon asymmetry. In Sec.~\ref{sec:results} we solve these phenomenological equations and identify regions of the parameter space leading to the correct baryon asymmetry. In Sec.~\ref{sec:model} we present a minimal model exhibiting a CP-violation in the cross-sections for $X$ and $\bar{X}$ scattering, and we show that this CP-violation is less efficient than the phenomenological parameterization adopted in Refs.~\cite{Dolgov:2020kqj,Dolgov:2021tsa} and in the previous sections. We draw our conclusions in Sec.~\ref{sec:conclusions}.

\section{The mechanism}\label{sec:mechanism}
The $X$ and $\overline{X}$ species evolve according to the interplay between different processes:  the absorption by the cosmological distribution of PBHs with number density $n_{\text{PBH}}$, the interaction with the relativistic plasma, and the decay via a process $X\to b+\text{SM}$, where $b$ is a SM particle with baryon number. We discuss these terms in turn.
\subsection{Absorption rate due to PBHs}
In Ref.~\cite{Dolgov:2020kqj}  the rate of infalling $X$ particles per unit time is estimated as
\begin{equation}
    \frac{dN_X}{dt}=\frac{4\pi M m_X n_X}{n_0 g \sigma_X T M^2_{\text{Pl}}},\label{rate}
\end{equation}
where $M_{\text{Pl}}$ is the Planck mass, $g\simeq 100$ is the number of relativistic interacting species in the plasma surrounding the PBH, $\sigma_X$ is the cross-section for scattering of $X$ particles with the SM, $n_0$ is the number density of each species in the plasma, $T$ is its temperature, and $n_X$ is the number density of $X$ particles. An equivalent expression can be obtained for $\overline{X}$. For the cross-section $\sigma_X$ we assume the unitarity bound $\sigma_X=f^4/m_X^2$. The cross-sections $\sigma_X$ and $\sigma_{\overline{X}}$ are slightly different due to CP violation. In this section and in Sec.~\ref{sec:results} we write the relative deviation in the cross-sections, which we will refer to as CP-violation efficiency, as
\begin{equation}
\sigma_X-\sigma_{\overline{X}}\simeq \varepsilon\left(\frac{m_X}{T}\right)f^2 \sigma_X.
\end{equation}
The functional form $\varepsilon(T)$ depends on the diagrams contributing to the asymmetry. In Refs.~\cite{Dolgov:2020kqj, Dolgov:2021tsa} $\varepsilon\left(\frac{m_X}{T}\right)$ is taken to be $1$. In Sec.~\ref{sec:model} we will derive, in the context of a specific field theoretical model, a different form for $\varepsilon\left(\frac{m_X}{T}\right)$ which is therefore theoretically motivated. With this choice, we will show that the feasibility of the mechanism may be spoiled in the context of a full model.

The absorption rate is determined by Eq.~(\ref{rate}) and is given by
\begin{equation}
    \left(\frac{dn_X}{dt}\right)_{\text{abs}}=-\frac{4\pi  m_X^3 \beta T_{\text{form}}}{gf^4 T M^2_{\text{Pl}}}n_X
\end{equation}
where 
$\beta=\rho_{\rm PBH}(T_{\text{form}})/\rho_{\gamma}(T_{\text{form}})$ denotes the ratio between the density of PBH  and the density of radiation at their formation at a temperature $T_{\text{form}}\simeq 3\cdot 10^{15}\;\text{GeV} (M_g)^{-1/2}$~\cite{masina2020dark}. 

The interaction with the plasma via the annihilation and creation processes $X\overline{X}\rightleftharpoons \text{SM}$ maintains thermal equilibrium before the species becomes non-relativistic. We assume that the coupling for this interaction is the same coupling $f$ as for the collision cross-section, as can be expected because of crossing symmetry. After the species becomes non-relativistic, the interaction rate rapidly drops and $X$ undergoes freeze-out, as we will verify explicitly in the solutions shown below. The rate of interaction with the plasma is
\begin{eqnarray}
    \left(\frac{dn_X}{dt}\right)_{\text{int}}
    -\frac{f^4}{m_X^2}(n_X n_{\overline{X}}-n_{\text{th}}^2),
\end{eqnarray}
where $n_\text{th}$ is the number density at thermal equilibrium.

The order of magnitude of the decay rate is $\Gamma=h^2 m_X$ where $h$ is the coupling of the $X$ decay process. We assume here that $X$ decays into much lighter particles, so that phase space factors are negligible. Furthermore, we assume that particles are at rest, since absorption becomes relevant for non-relativistic particles, and we solve the equations only in this regime. Then we have
\begin{equation}
    \left(\frac{dn_X}{dt}\right)_{\text{dec}}=-h^2 m_X n_X.
\end{equation}
\subsection{Evolution equation}
Collecting the interaction, absorption, and decay terms, we obtain the evolution equations as
\begin{eqnarray}
\frac{dN}{dx}&=&-\frac{\lambda}{x^2}(N^2-Y_{\text{th}}^2)-\alpha x^2 N-\mu x N,\label{eq:Nevolution}\\
\frac{dA}{dx}&=&\alpha x^2 (\varepsilon(x)f^2 N-A)-\mu x A,\label{eq:asymmetryevolution}
\end{eqnarray}
where we have introduced the yields $N\equiv (n_X+n_{\overline{X}})/2s$, $A \equiv (n_X-n_{\overline{X}})/s$ and $x \equiv m_X/T$. Furthermore, we introduce the dimensionless parameters 
\begin{equation}
\begin{split}
   \lambda&=\sqrt{\frac{\pi}{45}}  \frac{f^4 \sqrt{g_*} M_{\text{Pl}} }{m_X},\\
   \alpha&=\frac{ \pi^{7/2}}{ \sqrt{5} \zeta(3)} \, \frac{ \beta T_{\text{form}}}{g f^4\sqrt{g_*} M_{\text{Pl}}},\\
   \mu&=\sqrt{\frac{45}{4\pi^3}} \frac{h^2 M_{\text{Pl}}}{\sqrt{ g_*}m_X},
   \end{split}
\end{equation}
and $Y_{\text{th}}$ is the equilibrium yield for the species $X$. The quantity $g_*\simeq 100$ is the number of relativistic d.o.f. appearing in the entropy density.
\subsection{Baryon asymmetry}
Assuming that the only source of asymmetry is the decay of the $X$ particles we have
\begin{equation}
    \frac{d\eta^{\text{th}}_B}{dx}=\mu x A.\label{eq:etaevolution}
\end{equation}
\subsection{Constraints}
The mechanism is characterized by the following free parameters: the PBH mass $M$; the mass of $X$ particle, $m_X$; the two couplings $f$ and $h$; the fraction of PBHs at their formation $\beta$. 

The solutions obtained within this scenario are acceptable only if a number of physical constraints are verified: {\it i)} $\beta$ must be smaller than the known constraints, which are conveniently summarized in Ref.~\cite{Carr:2020gox} assuming a lighter dark matter candidate; {\it ii)} the entropy generated in the decay of the $X$ particles should not be as large as to dilute too much the generated asymmetry; {\it iii)} the whole process must take place much before Big Bang Nucleosynthesis; {\it iv)} $X$ decays after PBHs have evaporated, so that the asymmetry in the $X$ sector has reached its largest possible value; {\it v)} the annihilation process $X\overline{X}\to$SM induced by the coupling $h$ is negligible compared with the annihilation induced by the coupling $f$; {\it vi)} we determine the ratio between PBHs and radiation energy density at PBH evaporation as $\rho_{\text{PBH}}/\rho_{\text{rad}}\simeq \beta \, T_{\text{form}}/T_{\text{e}}$, and we verify that it is smaller than 1: this guarantees that the evolution proceeds under conditions of radiation domination; {\it vii)} we determine the accretion rate of mass from each PBH according to Eq.~\ref{rate} at $T_{\rm e}$, and we require that it is smaller than the evaporation rate, namely
     \begin{equation}\label{eq:acctoevaratio}
         R_{\text{ae}}=\left(\frac{dM}{dt}\right)_{\text{acc}}/\left(\frac{dM}{dt}\right)_{\text{eva}}<0.1,
     \end{equation}
     where $\left(\frac{dM}{dt}\right)_{\text{acc}}=m_X \frac{dN_X}{dt}$. 
     PBH evaporation would otherwise be delayed by the larger accretion rate until all the species had been completely absorbed.

\section{An explicit model realization of CP violation}\label{sec:model}
In the previous section we have outlined the mechanism in general. The one thing that remains to be specified is the form of the CP violation efficiency $\varepsilon(x)$. This must come from an analysis of the diagrams contributing to the CP violation, and may only be determined exactly in the context of a specific model. We remind the reader that Refs..~\cite{Dolgov:2020kqj,Dolgov:2021tsa} estimate this quantity to be equal to $1$. In this section, we propose a concrete model leading to a non-vanishing CP violation, and we estimate $\varepsilon(x)$ within this model.

We introduce as additional species to the Standard Model heavy fermion particles $X$, $Y$, and $b$, and two heavy scalar mediators $\phi$ and $\psi$, and we assume they couple to the Standard Model species $a$ and $c$ (which may be quarks, as a reference example) with the Lagrangian
\begin{eqnarray}
    \mathcal{L}_\mathrm{int}=-g_{\bar{a}X}\bar{\phi}\bar{a}X-g_{\bar{c}X}\bar{\phi}\bar{c}X \\ \nonumber-g_{\bar{b}Y}\bar{\phi}\bar{b}Y-g_{\bar{Y}X}\psi\bar{Y}X-g_{\bar{b}a}\psi\bar{b}a+\mathrm{h.c.}
\end{eqnarray}
All couplings are assumed to be complex numbers. The structure of the coupling can be justified if the new particles are charged under an additional conserved (but not necessarily gauged) $U(1)$, for example assuming that $X$ and $Y$ have charge $+1$, $\phi$ has charge $-1$, and $\psi$ is neutral. Notice that this symmetry would also allow the additional interaction terms $\mathcal{L}'_\mathrm{int}=-g_{\bar{b}c}\psi\bar{b}c$. The effect of this term is not essential to the model, and can be absorbed into a renormalization of the CP-violating efficiency, as we will see below. 

\begin{figure*}
    \centering
    \includegraphics[width=0.9\textwidth]{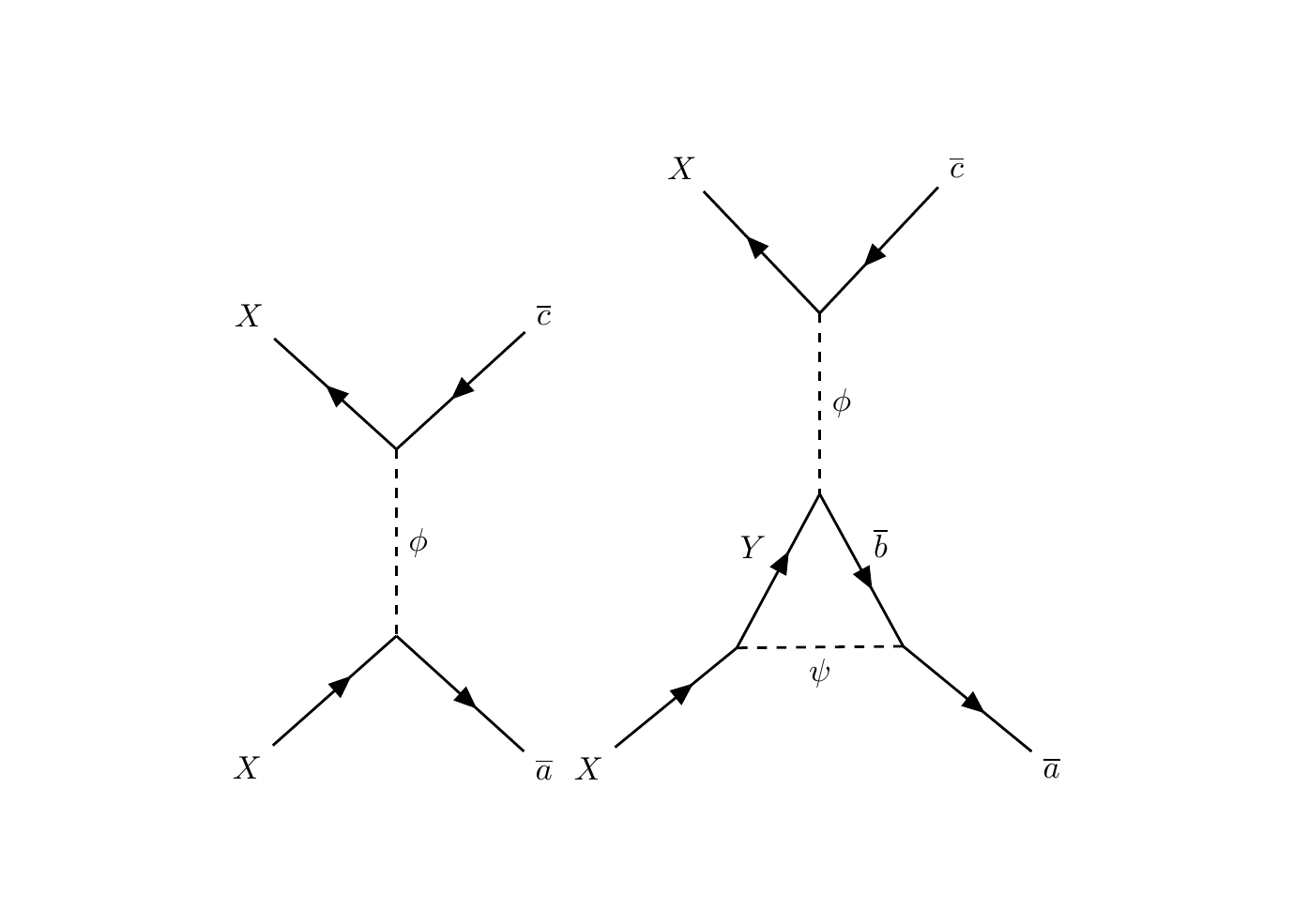}
    \caption{Feynman diagrams contributing to the scattering $X\bar{a}\to X\bar{c}$ to one-loop level.}
    \label{fig:feynman_diagrams}
\end{figure*}

Within this model, the cross-sections for scattering $X\bar{a}\to X\bar{c}$ and $\bar{X}a\to\bar{X} c$ are different at loop level, because of the interference between the diagrams in Fig.~\ref{fig:feynman_diagrams}. The additional coupling $g_{\bar{b}c}$ would lead to the appearance of a third diagram with a triangle loop in the final state. Thus, the CP violation would be renormalized by this additional term. Importantly, if the particle $c$ coincides with the particle $a$, the contributions from these two diagrams would not add up, as suggested in Ref.~\cite{Dolgov:2021tsa}, but rather it would cancel. This is a special case of a general result descending from CPT and Lorentz invariance: the cross-sections for $X\bar{a}\to X\bar{a}$ and $\bar{X}a\to\bar{X}a$ are necessarily equal.

The interference between the two diagrams in Fig.~\ref{fig:feynman_diagrams} creates a difference $\delta\sigma= \sigma_{X\bar{a}\to X\bar{c}}-\sigma_{\bar{X}a\to\bar{X}c}$
\begin{eqnarray}
    &\delta\sigma=-4\int \frac{d^3 \mathbf{p'}}{(2\pi)^3 2 E_{p'}}\frac{d^3 \mathbf{q'}}{(2\pi)^3 2 E_{q'}}\frac{1}{4E_p E_q}(2\pi)^4\\
    &\nonumber\delta^4(p'+q'-p-q)\mathrm{Im}\left(g^*_{\bar{a}X} g_{\bar{Y}X} g_{\bar{b}Y} g_{\bar{b}a}^*\right)\frac{|g_{\bar{c}X}|^2}{(s-m_\phi^2)^2}\mathrm{Re}\left[\ell\right],
\end{eqnarray}
where $s$ is the first Mandelstam invariant, $p$, $q$, $p'$, and $q'$ are respectively the four-momenta of the incoming $X$ and $\bar{a}$ particles and of the outgoing $X$ and $\bar{c}$ particles. For each of these four-momenta, we denote in bold the spatial 3-momentum and by $E_p$, $E_q$, $E_{p'}$, and $E_{q'}$ their energies in the laboratory frame. Finally, $\ell$ is the loop integral defined by
\begin{eqnarray}
    &\ell=\int\frac{d^4 k}{2(2\pi)^4}\mathrm{Tr}\left[(\slashed{p}'+m_X)(\slashed{q}'-m_c)\right]\\
    &\nonumber\frac{\mathrm{Tr}\left[(\slashed{k}-m_b)(\slashed{p}+\slashed{q}+\slashed{k}+m_Y)(\slashed{p}+m_X)(\slashed{q}-m_a)\right]}{(k^2-m_b^2)\left[(p+q+k)^2-m_Y^2\right]\left[(q+k)^2-m_\psi^2\right]},
\end{eqnarray}
where by $m_i$ we denote the mass of the $i$-th particle. We notice that $\ell$ depends only on the initial four-momenta $p$ and $q$, and not on the final four-momenta $p'$ and $q'$\footnote{The term $\mathrm{Tr}\left[(\slashed{p}'+m_X)(\slashed{q}'+m_c)\right]$ can be rearranged as $4(m_am_c+p\cdot q+\frac{m_a^2-m_c^2}{2})$, so it also does not depend on the final-state properties.}, so that it can be taken outside of the integral. Therefore, the relative deviation in the cross-section can be expressed to lowest order in the couplings as
\begin{eqnarray}
    &\frac{\delta\sigma}{\sigma_{X\bar{a}\to X\bar{c}}}=-4\frac{\mathrm{Im}\left(g^*_{\bar{a}X} g_{\bar{Y}X} g_{\bar{b}Y} g_{\bar{b}a}^*\right)}{|g_{\bar{a}X}|^2}\\
    \nonumber &\frac{2\mathrm{Re}\left[\ell\right]}{\mathrm{Tr}\left[(\slashed{p}'+m_X)(\slashed{q}'-m_c)\right]\mathrm{Tr}\left[(\slashed{p}+m_X)(\slashed{q}-m_a)\right]}.
\end{eqnarray}
The real part of the loop integral can be exactly determined using the Cutkosky rules, under the simplifying assumption that $m_\psi>m_X,m_a$. We will further simplify the result by assuming $m_a$ to be negligible, which is justified since the center-of-mass energies of the collisions in the primordial universe at the temperatures of interest are much higher than the masses of the quarks.  The final result is
\begin{eqnarray}
    \frac{\delta\sigma}{\sigma_{X\bar{a}\to X\bar{c}}}=\frac{\mathrm{Im}\left(g^*_{\bar{a}X} g_{\bar{Y}X} g_{\bar{b}Y} g_{\bar{b}a}^*\right)}{|g_{\bar{a}X}|^2} \frac{1}{4\pi s^2 m_\psi^2}\\
    \nonumber (s-(m_b-m_Y)m_X)(s-(m_b+m_Y)^2)\\
    \nonumber \sqrt{\left[s-(m_b+m_Y)^2\right]\left[s-(m_b-m_Y)^2\right]},
\end{eqnarray}
where $s$ is the Mandelstam invariant of the collision. 

At temperatures $T\ll m_X$, at which the absorption of particles onto primordial black holes becomes relevant, $s=m_X^2+2\frac{m_X T}{\sqrt{1-v^2}}(1-v\cos\theta)$, where $v$ is the velocity of $X$ and $\theta$ is the angle between the directions of $X$ and $\bar{a}$. Assuming $v\ll 1$, we expect $s\simeq m_X^2+2m_XT$ in order of magnitude. Furthermore, we choose $m_Y+m_b\simeq m_X$. The reason for this choice is that, if $m_b+m_Y<m_X$, the reactions $X\bar{a}\to Y\bar{b}$ would maintain chemical equilibrium for the $X$ species even after it becomes non-relativistic. $X$ would therefore rapidly disappear from the plasma because of the inhibition of inverse annihilation. Finally, we choose $|m_b-m_Y|\ll T$ and $m_\psi\simeq m_X$. These choices maximize the asymmetry in the cross-section, which becomes of the order
\begin{equation}\label{eq:asymmetryoom}
    \frac{\delta\sigma}{\sigma_{X\bar{a}\to X\bar{c}}}\simeq \frac{\mathrm{Im}\left(g^*_{\bar{a}X} g_{\bar{Y}X} g_{\bar{b}Y} g_{\bar{b}a}^*\right)}{|g_{\bar{a}X}|^2}\frac{1}{\sqrt{2}\pi}\left(\frac{T}{m_X}\right)^{3/2}.
\end{equation}

This expression contains the final result we aimed to obtain in this section, namely the temperature dependence of the CP violation. Expressed in the notation of Sec.~\ref{sec:mechanism}, we may write within this model
\begin{equation}
    \varepsilon(x)=\frac{1}{\sqrt{2}\pi x^{3/2}}.
\end{equation}

Compared with the assumption made in Refs.~\cite{Dolgov:2020kqj,Dolgov:2021tsa}, namely that $\varepsilon(x)=1$, here we find that accounting for a consistent particle model lowers the CP violation at lower temperatures, due to the factor $(T/m_X)^{3/2}$. This effect derives from the Cutkosky rules, which link the violation at loop level with the amplitude for scattering onto the intermediate state $X\bar{a}\to Y\bar{b}$: the latter clearly vanishes as $T\to 0$. In the next section, we will show the results which are obtained in this mechanism using both choices for $\varepsilon(x)$. 

\section{Results}\label{sec:results}

In this section we show our results for the asymmetry produced. For clarity, we divide our discussion according to the two choices outlined for the CP violation efficiency $\varepsilon(x)$: (i) the phenomenological scenario adopted in Refs.~\cite{Dolgov:2020kqj,Dolgov:2021tsa}, with $\varepsilon(x)=1$; (ii) the explicit model realization outlined in the previous section, with $\varepsilon(x)=\frac{1}{\sqrt{2}\pi x^{3/2}}$.
\subsection{Phenomenological scenario}

\begin{figure*}[ht]
\includegraphics[width=\textwidth]{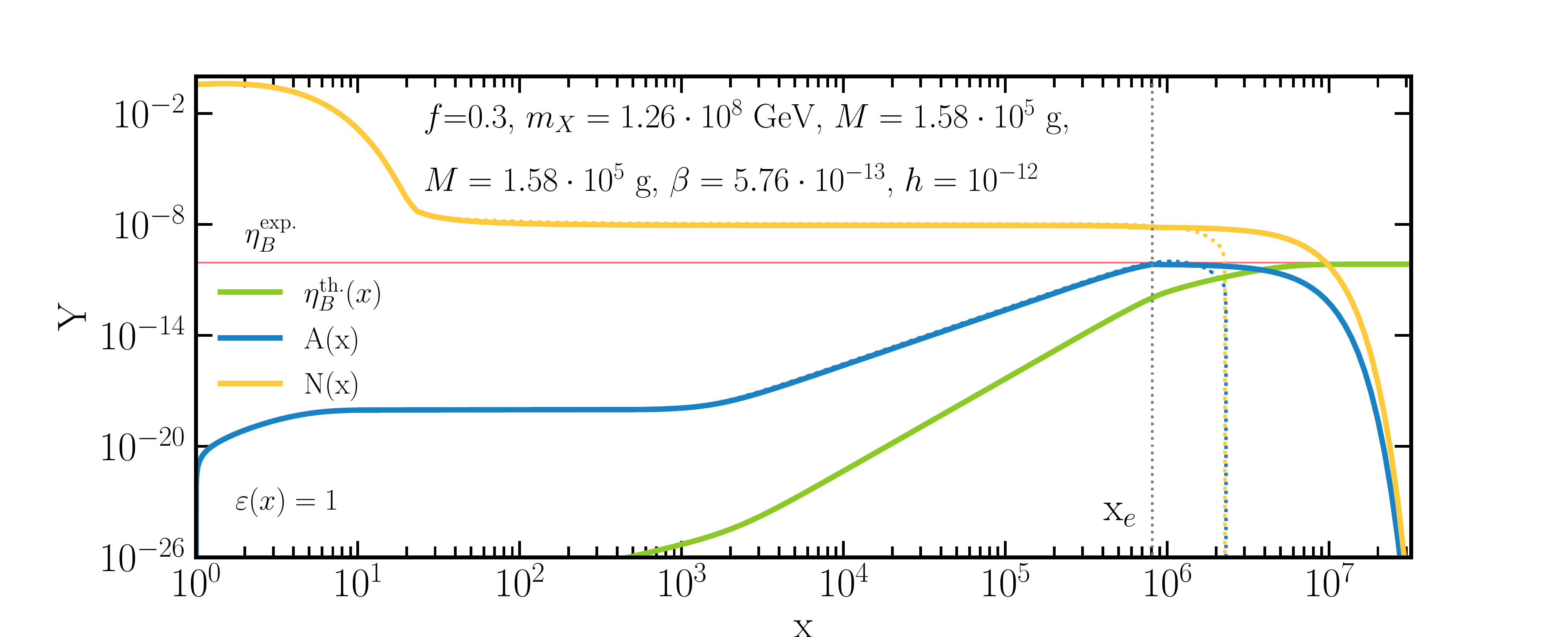}
\caption{Evolution of $X$ concentration (yellow), $X$ asymmetry (blue), and baryon asymmetry (green) as a function of $x$ for a benchmark choice of parameters. $\text{x}_e$ is the moment at which PBHs evaporate. The dotted curves represent the solution if PBH evaporation is not taken into account. We assume $\varepsilon(x)=1$ as in the phenomenological scenario. 
}
\label{Plot:benchmark}
\end{figure*}

The production of baryon asymmetry in both scenarios can be understood by exhibiting a benchmark solution to the above equations for a specific parameter choice. We show this solution for the phenomenological scenario with $\varepsilon(x)=1$ in Fig.~\ref{Plot:benchmark}.

The yellow curve, denoting the average yield $N$ for $X$ and $\overline{X}$, undergoes freeze-out at $x$ close to 10, reaching a constant value determined by the freeze-out process. Before freeze-out, the asymmetry (blue curve) increases and reaches a plateaux value, due to the initial rapid decrease in $N$; this is a first residual component of asymmetry. After freeze-out, $N$ remains approximately constant and undergoes a slow absorption by PBHs leading to a continuous production of asymmetry, represented by the steadily increasing blue curve; this is a second component of asymmetry which starts being produced after freeze-out. We emphasize that, even though the process of freeze-out happens at $x\sim10$, the continuous production of asymmetry dominates over the residual (plateaux) asymmetry only later, at $x\sim10^3$. The increase of the asymmetry can be essentially parameterized as $A(x)\propto x^3$. With increasing $x$, absorption becomes stronger and would eventually deplete all the species $X$ and the asymmetry produced. However, this situation is avoided because of PBH evaporation, which abruptly stops the absorption process and therefore maintains the produced asymmetry. To highlight this point, we show as a dashed line the solution determined without accounting for PBH evaporation: indeed the dashed curve vanishes for $x\sim2\times10^6$. On the other hand, the solid lines remains constant immediately after PBH evaporation, when asymmetry stops being produced. At this stage, the decay into baryons allows us to transfer the asymmetry from the $X$ sector into the baryonic one, represented by the green curve. The concentration of $X$ particles, and their asymmetry, completely disappear. The average energy density of $X$ and $\overline{X}$ is converted into radiation, without significantly heating the primordial plasma.

One point is especially important: in Ref.~\cite{Dolgov:2020kqj} the need for small couplings was emphasized. However, small couplings lead to very large absorption which suppresses the produced asymmetry before PBHs evaporate. Therefore, the opposite requirement must be met in order to produce the necessary asymmetry. Indeed, our benchmark solution leading to the correct asymmetry in this phenomenological scenario requires $f$ of order $0.1$. 
\subsection{Explicit model realization}

\begin{figure*}[ht]
\includegraphics[width=\textwidth]{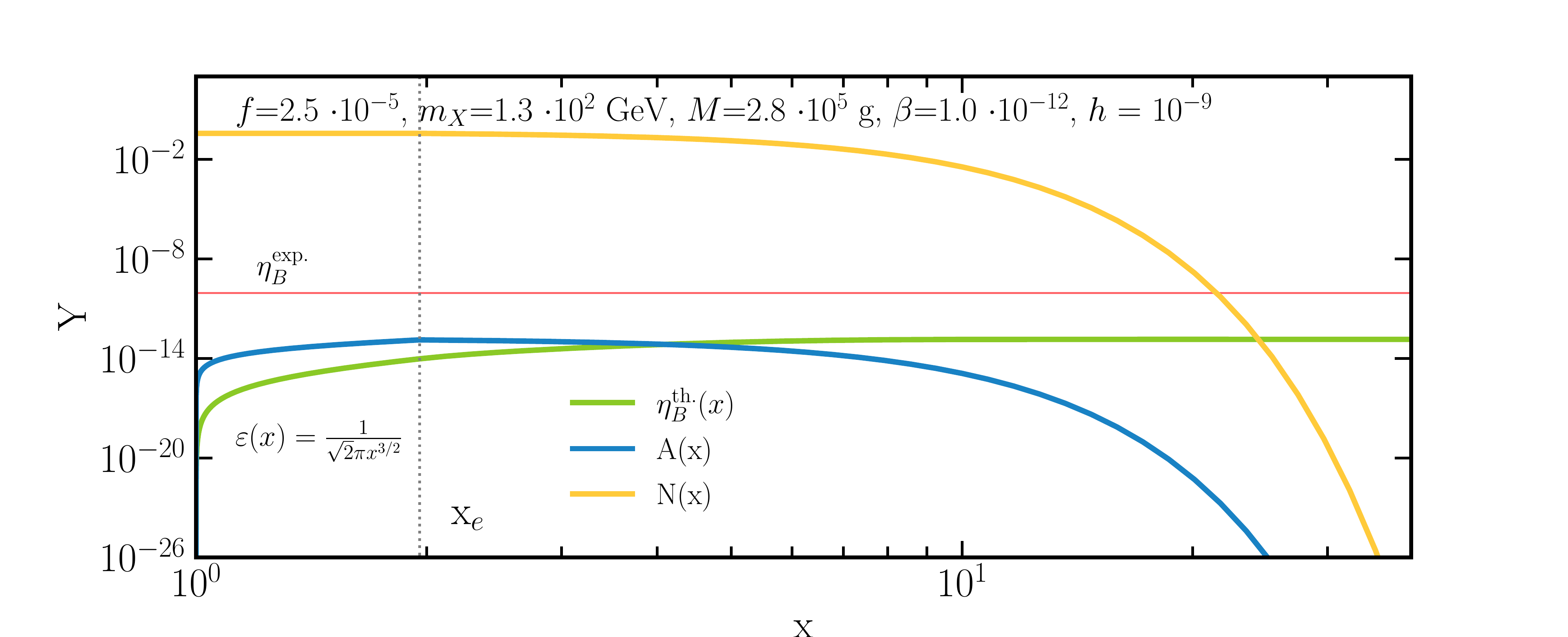}
\caption{Evolution of $X$ concentration (yellow), $X$ asymmetry (blue), and baryon asymmetry (green) as a function of $x$ for a benchmark choice of parameters. $\text{x}_e$ is the moment at which PBHs evaporate. We assume $\varepsilon(x)=\frac{1}{\sqrt{2}\pi x^{3/2}}$ as in the explicit model realization.
}
\label{fig:evolutionmodel}
\end{figure*}

In the model outlined in Sec.~\ref{sec:model}, the CP violation efficiency progressively decreases with time. Therefore, the growth of the asymmetry with $x$ is much slower than in the phenomenological scenario. This can be seen in Fig.~\ref{fig:evolutionmodel}, where we show the asymmetry evolution as a function of $x$ for a benchmark choice of parameters which maximizes the final asymmetry. In the region of $x<x_\mathrm{e}$, before PBHs evaporate, the asymmetry increases only as $A(x)\propto x^{3/2}$, in contrast with the phenomenological case where it grows as $A(x)\propto x^3$. For this reason, the final asymmetry produced is significantly lower than what could be attained in the phenomenological scenario. The maximal asymmetry that can be produced consistently with the constraints listed in Sec.~\ref{sec:mechanism} is of the order of $10^{-13}$, which is by three orders of magnitude lower than the experimental measurements.
We still reach the non-trivial result that this model can produce a baryon asymmetry. However, further study is required to develop a model quantitatively in agreement with the observed baryon asymmetry.

\section{Conclusions}\label{sec:conclusions}

The baryogenesis mechanism proposed in Refs.~\cite{Dolgov:2020kqj, Dolgov:2021tsa} works through the interplay between different factors: the freeze-out of a heavy species $X$, the CP-asymmetric absorption of this species by PBHs, and the subsequent decay of $X$, which both transfers the asymmetry to the baryonic sector and avoids the overproduction of non-relativistic $X$ particles which would otherwise dominate over radiation. Such a mechanism would provide an effectively new scenario for baryogenesis which does not need violation of the baryonic number at a Lagrangian level.
We have shown that, even at a phenomenological level, this scenario must be extended to account for PBH evaporation in order to produce the correct asymmetry. PBH evaporation is necessary to halt the absorption of the $X$ particles before they entirely disappear from the plasma, together with the asymmetry.

While this modification allows the mechanism to work at the phenomenological level, we have shown that it is still difficult to realize in an actual model. The scenario works only under the assumption of a CP violation efficiency independent of the temperature of the plasma, as suggested in Refs.~\cite{Dolgov:2020kqj, Dolgov:2021tsa}. However, the minimal model we were able to find leads to a CP violation efficiency progressively decreasing with decreasing temperature. This result seems to derive from the Cutkosky rules alone, and therefore seems to be robust. In this model, baryon asymmetry is indeed produced, but in a quantity which is too small compared with the experimental value. This suggests that a model quantitatively supporting the scenario is difficult to realize.

\vskip2.mm
{\it  Acknowledgments.}
We are grateful to Marco Chianese for useful comments. This work was supported by the Italian grant 2017W4HA7S, "NAT-NET" Neutrino and Astroparticle Theory Network (PRIN 2017), funded by the Italian Ministero dell’Istruzione, dell'Universit\`{a} e della Ricerca (MIUR), and Iniziativa Specifica TAsP of INFN.

\bibliographystyle{unsrt}
\bibliography{Database} 
\end{document}